\def\usepreprinttemplate{1} %
\newif\ifpreprint
\ifdefined\usepreprinttemplate
  \preprinttrue
\else
  \preprintfalse
\fi
\ifpreprint
  \documentclass[11pt]{article}
  \usepackage[a4paper, margin=0.9in, footskip=.6in]{geometry}
  \usepackage[T1]{fontenc}
  \usepackage{cochineal}
  \usepackage[cochineal,varg]{newtxmath}
  \usepackage{graphicx}
  \usepackage{url}
  \usepackage{placeins}
  \usepackage{caption}
  \usepackage{titlesec}
  \usepackage{titling}
  \usepackage[round,authoryear]{natbib}
  \usepackage[hidelinks]{hyperref}
  \titleformat{\section}{\normalfont\large}{\thesection}{1em}{}
  \titleformat{\subsection}{\normalfont\normalsize}{\thesubsection}{1em}{}
  \pretitle{\noindent\LARGE\bfseries\raggedright\sffamily}
  \posttitle{\par\vspace{0.5em}}
  \preauthor{\noindent\normalsize}
  \postauthor{\par\vspace{1.5em}}
  \predate{}
  \postdate{}
  \renewenvironment{abstract}{\noindent\textbf{Abstract.}\hspace{0.2em}\noindent}{\par\vspace{1em}\hrule height 0.4pt\vspace{1em}}
  \newcommand{\keywords}[1]{\par\medskip\noindent\textbf{Keywords: }#1}
  \newcommand{\titlerunning}[1]{}
  \newcommand{\authorrunning}[1]{}
  \newcommand{\institute}[1]{}
  \let\cite\citep
  \usepackage{array}
  \usepackage{booktabs}
  \usepackage{longtable}
\else
  \documentclass[runningheads]{llncs}
  \usepackage[T1]{fontenc}
  \usepackage{graphicx}
  \usepackage{url}
  \usepackage{placeins}
\fi

\graphicspath{{plots/}}

\title{Student Evaluation of Repeated AI Feedback Across a Semester of Writing}
\titlerunning{Student Evaluation of Repeated AI Feedback}

\ifpreprint
    \author{
   Andres Karjus\textsuperscript{1,2,3}\hspace{0.75em}
Janika Leoste\textsuperscript{1,4}\hspace{0.75em}
Tiia Õun\textsuperscript{1}\\
\textsuperscript{1}Tallinn University\hspace{0.75em}
\textsuperscript{2}Estonian Business School\hspace{0.75em}
\textsuperscript{3}University of Tartu\hspace{0.75em}
\textsuperscript{4}Tallinn University of Technology
    }
\else
    \author{/anonymized/}
    \authorrunning{}
    \institute{}
\fi

\date{\vspace{-1cm}}

\begin{document}
\maketitle

\begin{abstract}
Generative AI is increasingly used for feedback in higher education, but evidence from repeated classroom use remains limited. This short paper analyses 2988 reflective essay-feedback-appraisal instances from 283 Estonian bachelor students across one semester. Students obtained and assessed feedback from a self-selected AI tool using a uniform prompt. The present analysis of the anonymized text corpus covers essay content, AI feedback, and its perceived helpfulness. Students found feedback helpful and actionable more often than not; about a tenth thought AI unhelpful, more so towards the end of the semester. We also analyzed essay reflection depth, and used a validated AI text classifier to estimate the share of essays that could be treated as likely unaided student writing. The study contributes descriptive classroom evidence on integration of AI feedback --- a fast and scalable way to provide immediate writing advice, but not a self-contained route to better reflection. Benefits depend on whether students learn to use AI selectively and critically, without sliding into over-use harmful for the learning process.

\ifpreprint
\vspace{0.1cm}
\else
\fi
\end{abstract}

\section{Introduction}

AI-driven learning technologies are entering higher-education feedback workflows, including writing support, formative guidance, chatbot-mediated revision advice, and general-purpose tutoring \cite{dai_2023_can,escalante_2023_ai,lee_2024_harnessing,aru_2025_developing}. When aligned with the task and assessment design, such systems can achieve outcomes comparable to human feedback  \cite{kaliisa_2026_how}.
Students often value generative AI tools for rapid access and practical support, but also question their accuracy, contextual fit, tendency to produce generic advice, and risk of over-reliance \cite{chan_2023_students,zhang_2025_evaluating}.
Feedback literacy and evaluative judgment frame students as active interpreters who appraise, select, and decide how to use feedback \cite{carless_2018_development,tai_2018_developing}.
This evaluative work also involves trust calibration: learners must decide what to trust, what to verify, and when to reject AI advice \cite{lee_2004_trust,zhang_2025_evaluating}.
However, AI-feedback quality has often been assessed indirectly through short interventions and perception-based measures \cite{calli_2026_when}.
Reflective writing is one particularly demanding case as useful feedback should support self-observation, interpretation, and reconsideration of experience, not only correctness or structure \cite{hatton_1995_reflection,kember_2008_four}.
This observational study examines AI feedback in 13 iterations of students' reflective essays, connecting generated output to student appraisal rather than evaluating feedback in isolation or in retrospect, and asks:\\
RQ1: How do students appraise repeated AI feedback across assignments?\\
RQ2: How is feedback specificity associated with its helpfulness and quality?\\
RQ3: Is expressed helpfulness associated with reflection depth in the next essay?\\
RQ4: What is the share of essays likely written (not generated) by students, and how does estimated authorship relate to estimated depth?

\section{Methods}

The corpus comes from an Estonian university course on digital skills and AI for first-year humanities bachelor students in autumn semester 2025. It consists of 2988 homework (HW) triplets from 283 students: the 300--350 word essay on the weekly lecture, paragraph-length AI feedback from a self-selected AI chatbot using a shared course prompt, and the student's 1--2 sentence appraisal of whether the feedback was substantive and useful. The corpus was filtered from an initial total of 3508 HW submissions by all 299 students to include only complete triplets. The prompt was slightly revised after HW4 with instructions to give more concrete essay-grounded feedback and guidance for deeper reflection (this is taken into account in the analyses below). Human teaching assistants graded the full HW package on a 0--3 point scale (totaling 39\% of the final course grade). Full credit was attainable when students met the assignment requirements and included the required components. Grading considered content quality, but reflective depth itself was not a separate rubric dimension. 
The students were asked to report their AI service or chatbot and share a chat thread URL every time (7\% of parsed submissions nevertheless did not contain a URL). We can therefore quantify platform usage, but exact model versions, settings, account tiers, and system prompts were unobserved. 
AI usage in writing was never expressly prohibited; rather, the course content itself frequently discussed how and why AI-overuse and over-reliance can harm cognition and learning outcomes.
The course description stated that submitted coursework could be used for scientific purposes and offered an opt-out; none was received. The present study is a retrospective, fully anonymized secondary analysis of ordinary coursework, treated essentially as a text corpus; we only report aggregate results without identifying students or publishing text examples.

We used GPT-5.4 (via its responses API) for LLM-assisted annotation \cite{karjus_machine-assisted_2025} of the anonymized text corpus for four separate descriptive constructs: appraisal theme (7 categories), appraisal helpfulness stance (helpful, mixed, unhelpful), essay reflection depth using a four-category Hatton--Smith derived scheme \cite{hatton_1995_reflection,kember_2008_four}, and essay-groundedness or specificity (1--5 scale) of the AI feedback. The output labels are treated as proxies rather than ground truth. We use descriptive analyses and mixed effects models to explore the questions; coefficient estimates and broader outcome models are reported in the supplement.
Two of the coauthors conducted human validation on a set of 25 appraisals and essays; GPT any-human-match accuracy was 96\% for appraisal theme (human agreement 76\%, $\kappa=.64$), 96\% for helpfulness (human 88\%, $\kappa=.78$), and 84\% for essay depth (64\%, ordinal $\rho=.56$), suggesting adequate fit for descriptive use.

\ifpreprint
\begin{figure}[h]
\centering
\includegraphics[width=\textwidth]{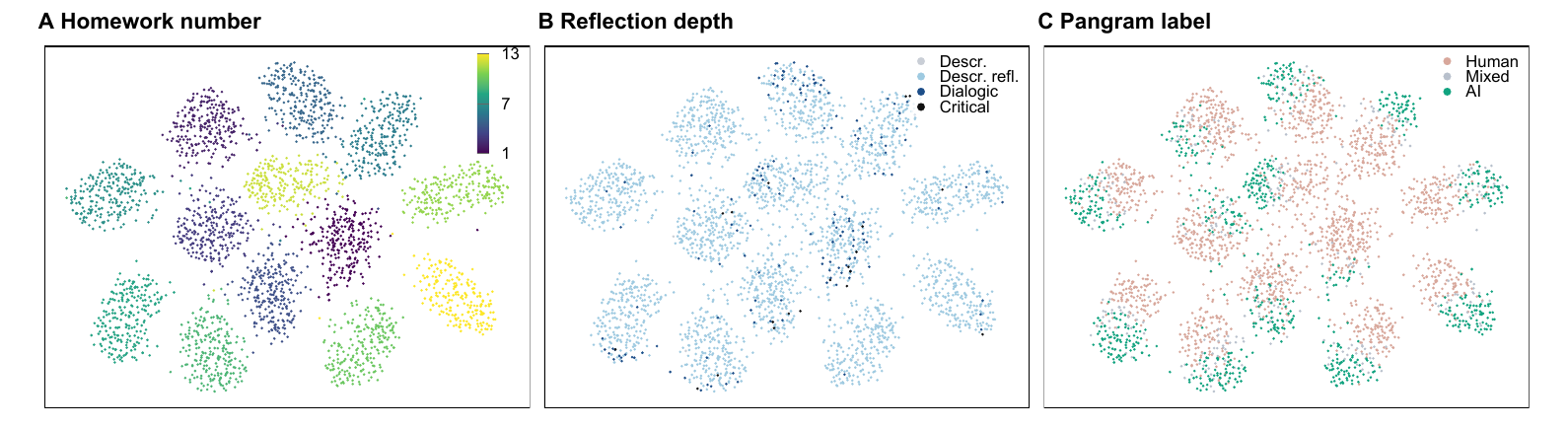}
\caption{
Dataset overview. Each essay text is a point, mapped onto a UMAP-projection of Fasttext embeddings (via doc2vec); proximity indicates semantic similarity, and panel A shows neat grouping of homeworks. Panel B shows where the deeper reflection texts are, and C colors text by predicted authorship, reflecting lexico-semantic differences in likely human vs likely AI-generated texts.
}
\label{fig_umaps}
\end{figure}
\fi

We also applied Pangram's AI-text detector (API version 3.3.2) to the essay texts as an aggregate text-classification check (this was only used retrospectively on the anonymized corpus, never as a basis for grading or student-level judgment). As Pangram does not list Estonian as an officially supported language, we evaluated it on close controls: 25 pre-2022 human reflection/blog texts about digital competence scraped from the open web (structurally and thematically similar to our essay corpus), and for contrast a set of 25 GPT-5.5-generated reflections matched to assignment genre and topics, and a hybrid set where human texts were lightly edited by GPT-5.5 to improve wording and conclusions. The detector separated the human and fully generated controls with 100\% accuracy; with no observed control errors, we did not estimate calibration intervals, and corpus labels are used only in aggregate. Of the 25 lightly edited texts, only 1 was labeled as mixed; mixed AI use is continuous, and validating graded AI assistance would require edit-distance or detector-score checks on a larger sample. We expect that classification labels can nevertheless err in both directions, as human prose may resemble AI-generated prose by chance (or learned stylistic convergence due to prolonged AI use), while generated but well-prompted prose may be labeled human. Detector labels here are descriptive estimates of AI-like essay submissions, not ground truth about individual authorship.

\begin{figure}[h]
\centering
\includegraphics[width=\textwidth]{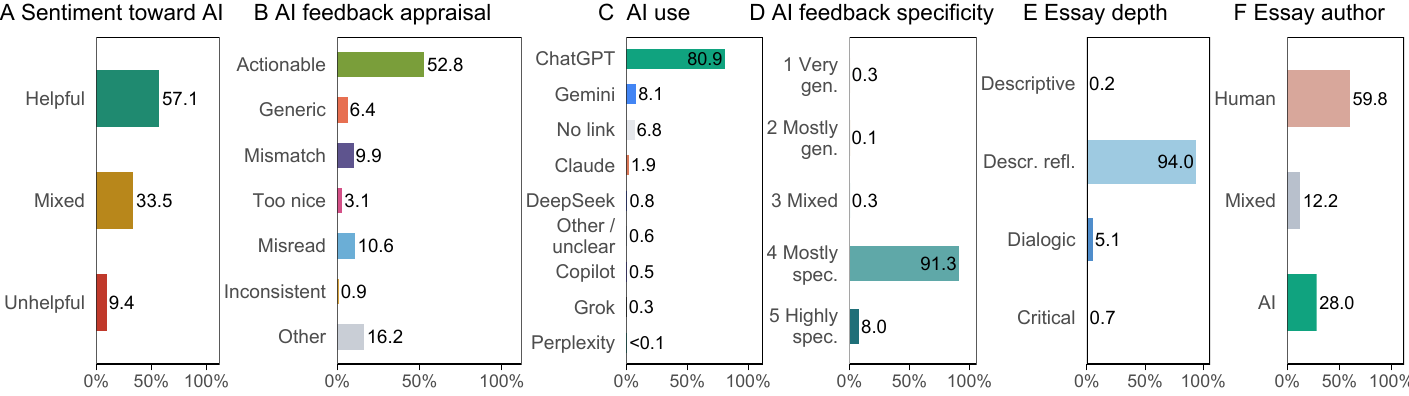}
\caption{Aggregate annotated distributions: expressed student sentiment toward AI feedback, appraisal theme, URL-inferred AI tool, estimated AI-feedback specificity, essay reflection depth, and Pangram-estimated essay-authorship label shares.}
\label{fig_summary_bars}
\end{figure}

\section{Results}

\begin{figure}[h]
\centering
\includegraphics[width=\textwidth]{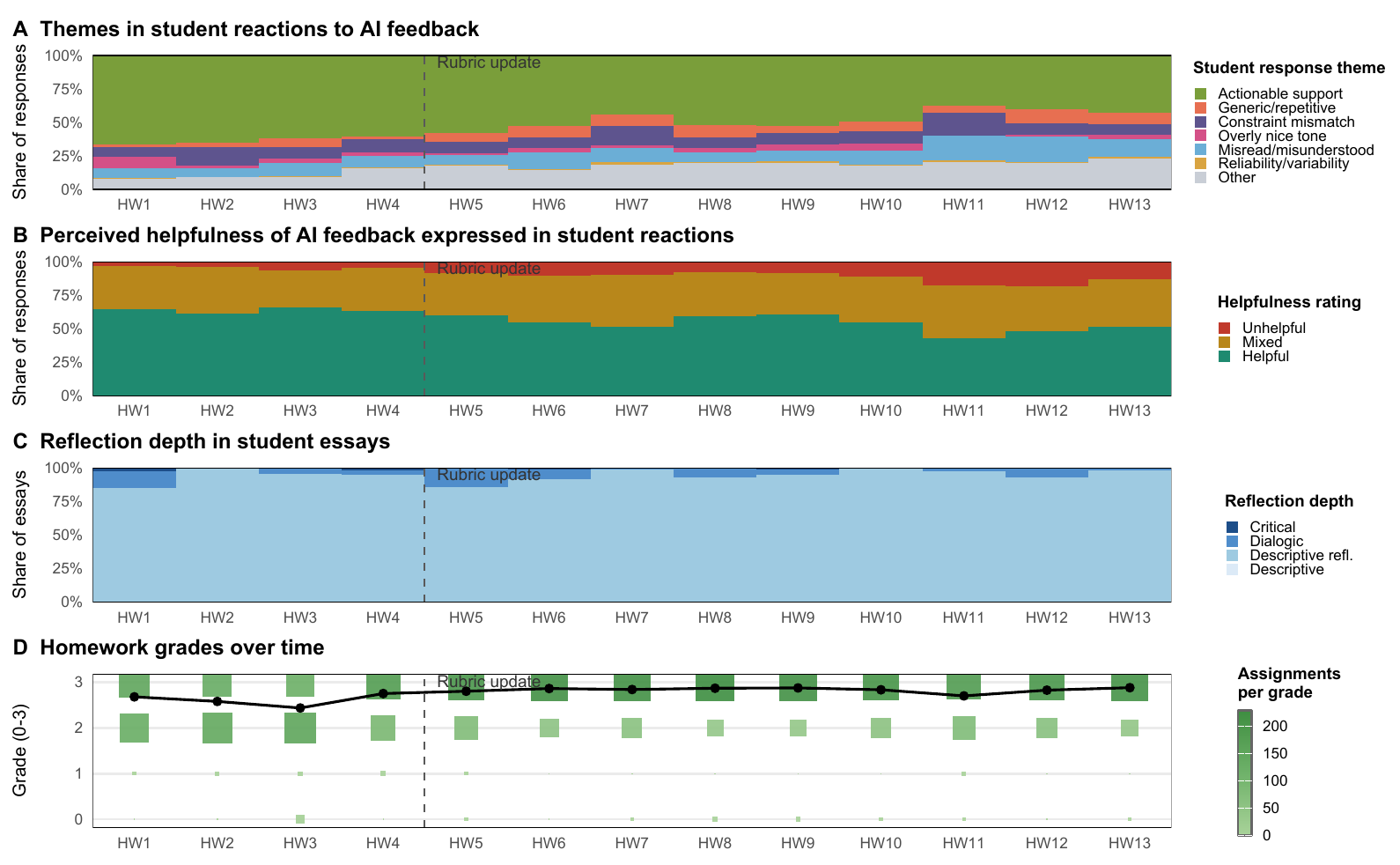}
\caption{Over-time distributions across homework 1--13: response theme, expressed helpfulness, essay reflection depth, and human-assigned grade points (square size corresponds to number of grades; line marks average). The dashed vertical line marks the switch to the second feedback-prompt version.}
\label{fig_over_time_panels}
\end{figure}

 Most students (94\%) wrote essays at a descriptive reflection level, and most used ChatGPT for feedback (81\%). AI produced specific feedback (99\% 4 or 5 on a scale of 1--5), which over half the students found helpful (57\%; 34\% mixed, 9\% unhelpful, answering our first RQ; see Fig. \ref{fig_summary_bars}). Across 2988 annotated appraisals, the most common theme was actionable support, meaning that students described AI feedback as giving usable suggestions or a clear next step (53\%). Both positive stances remained prevalent but fell over the semester, actionable-support from 67\% in HW1 to 43\% in HW13 (Fig. \ref{fig_over_time_panels}). Responses noting that the AI misread the essay or misunderstood a task detail peaked at 19\% in HW12. ChatGPT during that period served GPT-5 and later 5.1 generation models as defaults; Gemini, the second most common tool (8\%), defaulted to its 2.5 models.

Essay-depth coding was more stable than feedback-appraisal patterns. Most coded essays were descriptive reflection, meaning that they connected course content to personal experience but rarely moved into sustained alternative perspectives or critique. Dialogic reflection occurred in 5\% of essays, and the deepest category, critical reflection, in less than 1\%. Homework grading scores were compressed near the top of the 0--3 scale ($M$ = 2.77, $SD$ = 0.43), with 70\% at the maximum score. An ordinal mixed model showed no association between expressed helpfulness and next observed reflection depth after controlling current depth, position, and prompt version (RQ3).

We do not endorse AI detectors for grading or other pedagogical decisions about individual students, but used Pangram here as an aggregate text classifier because generated essays would change how later feedback appraisals should be interpreted. Pangram labeled 60\% of the essay corpus as likely human-written, 28\% as likely generated, and 12\% as mixed (totaling 40\% for some signal of AI use for a task where unaided student writing was encouraged; RQ4). The share classified as likely AI by Pangram rose from 8.5\% in HW1 to 40.3\% in HW13, a 4.8-fold increase.

\begin{figure}
\centering
\includegraphics[width=\textwidth]{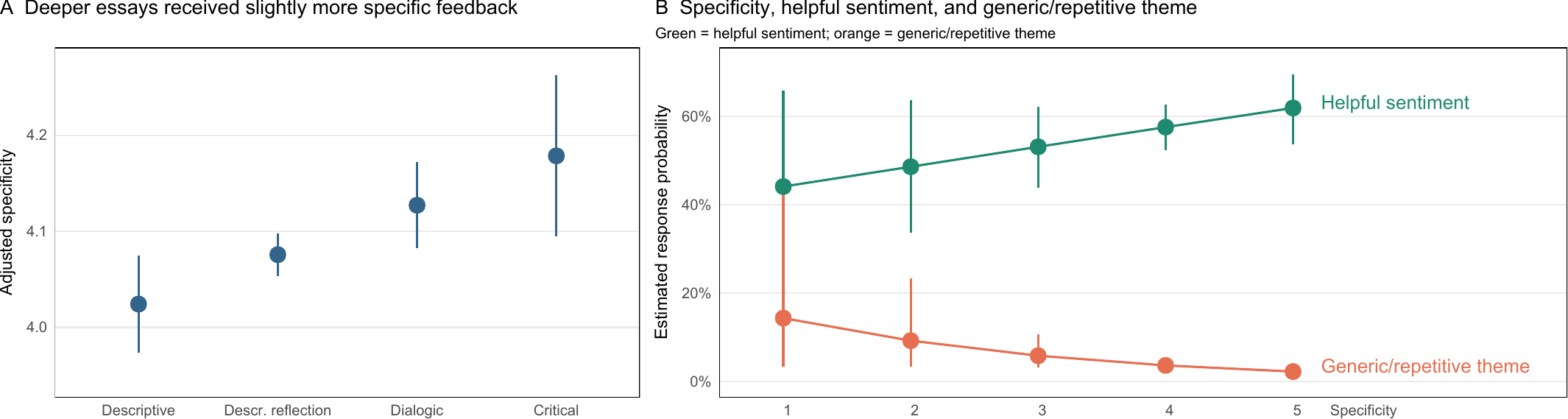}
\caption{Adjusted specificity-related estimates. Panel A shows estimated feedback specificity, operationalized as essay-groundedness, across essay-depth levels. Panel B shows predicted probabilities that the student response was coded as helpful or complained that feedback was generic or repetitive. Points are fitted values; bars show 95\% confidence intervals.}
\label{fig_feedback_specificity_effects}
\end{figure}

Feedback essay-groundedness was concentrated in the upper part of the specificity scale: 91\% of coded feedback texts were rated mostly specific, 8\% highly specific, and fewer than 1\% fell in the lower 1--3 range. Because 99\% of cases fell in the top two categories, this measure should not be interpreted as general feedback quality, actionability, or pedagogical value. After the prompt revision, lower-specificity cases fell from 2\% to less than 1\%, while score-5 feedback rose from 2\% to 11\%. Mixed-effects logistic models with student random intercepts and HW number confirmed both shifts: the revised prompt predicted lower odds of lower-specificity feedback (OR = 0.04, $p$ = .003) and higher odds of score-5 specificity (OR = 2.97, $p$ < .001). 
This mainly reduced complaints that feedback was generic or repetitive (RQ2). Groundedness helped reduce a sense of boilerplate, but was not sufficient by itself to make feedback feel useful (Fig. \ref{fig_feedback_specificity_effects}).

\FloatBarrier

\section{Discussion}

In this semester-long dataset, student appraisals shifted from generally accepting toward more differentiated evaluation.
Actionable support remained the most common theme, but later appraisals more often identified shortcomings such as generic advice, mismatch with the essay or task, and misreadings.
This pattern fits feedback-literacy and evaluative-judgment accounts: students were not only receiving feedback, but likely judging whether it was relevant, trustworthy, and usable \cite{carless_2018_development,tai_2018_developing}.
Because the feedback came from automated systems, this judgment also involved trust calibration: students had to decide when generated advice should be accepted, verified, or rejected \cite{lee_2004_trust,zhang_2025_evaluating}.
We captured cohort-level appraisal of AI feedback, although not whether individual students became better feedback evaluators over time.

The essay-groundedness results qualify what prompt-level grounding can and cannot accomplish. More specific feedback addressed one source of dissatisfaction about generic advice, but it did not settle judgments about accuracy, tone, or actionability. This reflects work showing AI-supported learning depends on course design, assessment expectations, learner agency, cognitive engagement, and teacher facilitation, not only on tool access or feedback quality \cite{leoste_2025_integration,kaver_2026_integration,laak_2025_ai}.
The unchanged descriptive reflection-depth result and relatively good grades should be read in light of the assessment design. Full credit was fairly easy to obtain due to a completion-oriented rubric. This does not necessarily mean that AI feedback had no educational value, but it is unlikely to deepen reflective writing by itself when the assignment does not directly incentivize it or require revision following feedback, comparison with explicit reflection criteria, follow-up questions, or a rationale for accepting or rejecting the given feedback. Such requirements would more likely keep learner responsibility visible in the dialogue with AI \cite{leoste_2025_socratic,gu_2026_fostering}.

The authorship estimates complicate the interpretation. Besides a third being classified as likely generated, 51\% of the relatively few essays coded as dialogic or critical reflection were likely generated. This does not mean generated texts produced deeper learning: such text can reproduce surface features of reflection, receive plausible feedback, and be trivially followed by a brief accepting appraisal --- while the student does little or none of the intended writing or evaluative work. We do not endorse classroom usage of (especially uncalibrated) AI detectors to judge individual students, but as a corpus analysis, the pattern points to a design problem for AI-integrated coursework. Comparable concerns have been raised in related fields like programming education: students may use AI for homework without reporting, leading to over-reliance and assessment-design risks \cite{leoste_2025_integration,kaver_2026_integration}. When a repeated low-stakes task legitimizes AI for one component, students may extend that delegation to the parts meant to exercise their own judgment. 
No single design feature completely prevents this. Prohibition can be ignored, while prescribed rationales, revisions, and self-reports can also be generated. Written tasks, especially those intentionally involving AI, would likely benefit from making the intended learning work less dependent on the final text alone, for example by tying writing and feedback use to earlier in-class writing, version histories, or brief peer- or teacher-facing explanations. This does not certify authorship, but shifts the task toward showing situated decisions around writing and feedback use, and away from producing a plausible submission. 

Several limits follow from the design and measurement strategy. This study comes from one course and one assignment format; students chose their own chatbots; provider labels come from URLs rather than exact model settings; the appraisal slot was short; LLM-assisted coding remains an approximation; and the authorship analysis indicates that a large share of the data likely cannot be treated as purely unaided student writing (while the classifier does not officially support Estonian, and our test set used to assess it was quite small, albeit specific). We also cannot ascertain whether increasing negativity and AI overuse were driven more by general semester fatigue or dissatisfaction with the task or course. 
Future work could combine repeated classroom data with experiments, detailed writing-process traces, and direct measures of feedback literacy and student motivation.

\section{Conclusions}

This semester-long study of 2988 homework items shows that integrating AI feedback into regular coursework adds more than a new source of advice; it can change what the writing task becomes. Students generally reported chatbot feedback as useful, especially when it offered concrete grounded suggestions, but over time their responses became more attentive to problems such as generic advice, content mismatch, or overly soft praise. At the same time, most essays remained descriptively reflective rather than moving toward deeper engagement, and a retrospective analysis of the anonymized corpus suggested that many submitted essays were likely AI-generated, more so by the end of the semester. AI feedback should therefore not be evaluated only by whether students like it, or whether the task's textual components appear plausible. Its educational value depends on the surrounding task: whether students still have to write, and are motivated to interpret, question, and integrate feedback as part of their own writing process.

\ifpreprint
\hrulefill
\paragraph{Author contributions.}
A.K. collected the data, analyzed the data, created the figures and wrote the paper. J.L. and T.Õ. annotated the human test sets, provided comments, and wrote parts of the paper.
\vspace{-0.8\baselineskip}
\paragraph{Funding.}
This work was partly supported by the European Union’s Horizon Europe research and innovation programme under grant agreement No 101119689 (EdTech Talents).
A.~Karjus was further supported by the European Union through the European Regional Development Fund under the Sectoral Mobility measure (SekMo), project number 2021-2027.1.01.25-1349. 
\vspace{-0.8\baselineskip}
\paragraph{Data availability.} Due to the nature of the corpus, it cannot be published as open data. The anonymized metadata and secondary annotations are available upon reasonable request from the corresponding first author.

\bibliographystyle{apalike}
\bibliography{lit}

\clearpage
\appendix
\setcounter{table}{0}
\renewcommand{\thetable}{A\arabic{table}}

\newcolumntype{P}[1]{>{\raggedright\arraybackslash}p{#1}}
\newcolumntype{R}[1]{>{\raggedleft\arraybackslash}p{#1}}
\newcommand{\PromptCell}[1]{\parbox[t]{0.47\textwidth}{\setlength{\parskip}{0pt}\setlength{\parindent}{0pt}\ttfamily\scriptsize\raggedright #1}}
\newcommand{\SinglePrompt}[1]{{\setlength{\parskip}{0pt}\setlength{\parindent}{0pt}\ttfamily\tiny\raggedright #1\par}}

\section*{Appendix}
\addcontentsline{toc}{section}{Appendix}

This appendix documents the parsing logic, the initial and updated student-facing feedback prompts in Estonian and English translation, the task-specific LLM annotation prompts, and supplementary model estimates.
No student text or personal data is reproduced here.

\section{Corpus parsing logic}

Submission HTML was converted to plain text and trailing chat-share link blocks were removed before segmentation.
Two candidate splitters recognized either repeated dash, underscore, or asterisk separators (including Unicode dash variants) or numbered markers for the second and third components (students were asked to follow the provided template, but many had replaced delimiters with their own symbols).
Empty fragments, preambles, headings, and trailing link-only lines were removed; surplus fragments after the third component were merged and flagged.
The higher-scoring candidate was retained, favoring exactly three components and plausible text lengths.
Rows were flagged when feedback or the student response was missing, feedback contained fewer than 40 characters, the response contained fewer than 12 characters, or surplus fragments had been merged; flagged rows were excluded from analyses requiring a complete essay--feedback--response triple.

\section{Student-facing feedback prompts}

\begin{longtable}{P{0.47\textwidth}P{0.47\textwidth}}
\caption{Initial student-facing feedback prompt.}\label{tab:prompt-v1}\\
\toprule
\textbf{Original Estonian} & \textbf{English translation} \\
\midrule
\endfirsthead
\toprule
\textbf{Original Estonian} & \textbf{English translation} \\
\midrule
\endhead
\bottomrule
\endfoot
\PromptCell{\# Sinu funktsioon ja üldine juhend\par
Sa oled aine ``Digipädevus ja tehisaru'' AI-abiõppejõud, sinu ülesandeks on tagasisidestada allpool esitatud üliõpilase enesereflektsiooni esseed (algab pärast kolme \texttt{"""} märki). Võta arvesse seda tagasisidestamise juhendit, arvesta kõigi neid aspektidega ja sünteesi selle põhjal üliõpilasele lühike ja konkreetne tagasiside (üks ilus kompaktne tekstilõik).\par
Essee teemaks on \texttt{[LEKTOR]} loengu \texttt{[PEALKIRI]} sisu (võimalik, et kirjutaja mainib ka lisaks loetud tekste või annab tagasisidet loengupidajale).}
&
\PromptCell{\# Your role and general instruction\par
You are the AI teaching assistant for the course ``Digital Competence and Artificial Intelligence''. Your task is to give feedback on the student's self-reflection essay presented below (it begins after the three \texttt{"""} marks). Take this feedback guide into account, consider all of its aspects, and on that basis synthesize brief and concrete feedback for the student (one polished, compact paragraph).\par
The essay's topic is the content of \texttt{[LECTURER]}'s lecture \texttt{[TITLE]} (the writer may also mention additional texts they have read or give feedback to the lecturer).}
\\
\PromptCell{\# Tagasisidestamise juhend: mida tagasisides mainida\par
\#\# Vorm ja keel, kontrolli kas:\par
- tekst on pikkuses umbes 300--350 sõna? (hinda suurusjärku)\par
- kirjutaja keelekasutus on üldiselt akadeemiline, selge, ja väldib üldsõnalisust?\par
- tekstil on loogiline struktuur (sissejuhatus-arutlus-kokkuvõte)?}
&
\PromptCell{\# Feedback guide: what to mention in the feedback\par
\#\# Form and language, check whether:\par
- is the text about 300--350 words long? (estimate roughly)\par
- is the writer's language generally academic and clear, and does it avoid vague or overly general wording?\par
- does the text have a logical structure (introduction--discussion--conclusion)?}
\\
\PromptCell{\#\# Sisu ja ülesande täitmine, analüüsi:\par
- kas essees on lühike ülevaade loengust ja mis mõtted kirjutajal tekkisid?\par
- aga kas essees väljendub lisaks kirjeldusele ka kuidas õppimine toimus ja kas/miks see kirjutaja jaoks oluline oli (mõtle tagasisidestades Ash\&Clayton DEAL mudelile, aga seda nimeliselt tsiteerida pole vaja)\par
- kas kirjutaja on loonud isiklikud seosed: loengu sisu varasemate teadmiste või väärtuste või hoiakute kontekstis; ja kas kirjutaja mainib, kas/kuidas õpitut saaks kasutada, kas on midagi mida teisiti teha või mõelda? (vastus võib ka olla, et hoiakud ei muutunud või uusi teadmisi ei saanud)\par
- ja kokkuvõtteks, kas kirjutaja piirdub vaid kirjeldamisega või liigub vähemalt ühe sammu võrra sügavamale (nt seletab, küsib, analüüsib, seob laiemalt mõne teksti või ühiskondliku nähtusega). Analüüsi, mis tasemel refleksioon asub ja suuna vajadusel järgmine kord minema sügavamale (lähtu tagasisidestades Hatton\&Smith nelja tasandi süsteemist, aga seda nimeliselt tsiteerida pole vaja).}
&
\PromptCell{\#\# Content and task completion, analyze whether:\par
- does the essay contain a brief overview of the lecture and the thoughts that arose for the writer?\par
- in addition to description, does the essay also express how learning took place and whether/why it was important to the writer? (When giving feedback, think of the Ash\&Clayton DEAL model, but there is no need to cite it by name.)\par
- has the writer made personal connections, placing the lecture content in the context of prior knowledge, values, or attitudes; and does the writer mention whether/how what was learned could be used, or whether there is anything to do or think differently? (The answer may also be that attitudes did not change or no new knowledge was gained.)\par
- finally, does the writer remain at description or move at least one step deeper (for example, explain, question, analyze, or connect more broadly to another text or social phenomenon)? Analyze the level of reflection and, if needed, guide the writer to go deeper next time (base the feedback on Hatton\&Smith's four-level system, but do not cite it by name).}
\\
\PromptCell{\#\# Tagasiside toon ja vorm:\par
- ole toetav aga konstruktiivelt kriitiline! (isegi kui essee on puudulik, püüa alguses võimalusel leida vähemalt mõni positiivne külg, enne kui kritiseerid)\par
- too välja, millised reflektsiooni aspektid on täidetud hästi ja milliseid (kui on puudulikud) võiks edaspidi arendada!\par
- seosta soovitused kirjutaja enda tekstis toodud mõtetega; soovitusi andes väldi üldsõnalisust!\par
- ole väga konkreetne ja ära raiska sõnu; tagasiside peaks olema üks lõik.\par
- lõpus esita avatud küsimus (``Edasi mõtlemiseks''), mis julgustab kirjutajat mingil teemal või suunal edasi mõtlema (aga ära esita sellist küsimust, millele kirjutaja on juba tegelikult vastanud).}
&
\PromptCell{\#\# Tone and form of the feedback:\par
- be supportive but constructively critical (even if the essay is deficient, try, if possible, to find at least one positive aspect at the beginning before criticizing)\par
- point out which aspects of reflection are done well and which ones, if deficient, could be developed further\par
- connect the suggestions to the ideas presented in the writer's own text; avoid vague or overly general suggestions\par
- be very concrete and do not waste words; the feedback should be one paragraph\par
- at the end, ask an open question (``For further thought'') that encourages the writer to think further about a topic or in a particular direction (but do not ask a question that the writer has already actually answered)}
\\
\PromptCell{\# Anna nüüd tagasisidet sellele esseele:\par
\texttt{"""}\par
\texttt{[lisa siia oma tekst]}}
&
\PromptCell{\# Now provide feedback on this essay:\par
\texttt{"""}\par
\texttt{[insert the essay text here]}}
\end{longtable}

\begin{longtable}{P{0.47\textwidth}P{0.47\textwidth}}
\caption{Updated student-facing feedback prompt.}\label{tab:prompt-v2}\\
\toprule
\textbf{Original Estonian} & \textbf{English translation} \\
\midrule
\endfirsthead
\toprule
\textbf{Original Estonian} & \textbf{English translation} \\
\midrule
\endhead
\bottomrule
\endfoot
\PromptCell{\# Sinu funktsioon ja üldine juhend\par
Sa oled aine ``Digipädevus ja tehisaru'' AI-abiõppejõud, sinu ülesandeks on tagasisidestada allpool esitatud üliõpilase enesereflektsiooni esseed (algab pärast kolme \texttt{"""} märki). Võta arvesse seda tagasisidestamise juhendit, arvesta kõigi neid aspektidega ja sünteesi selle põhjal üliõpilasele lühike ja konkreetne tagasiside (üks ilus kompaktne tekstilõik).\par
Essee teemaks on \texttt{[LEKTOR]} loengu \texttt{[PEALKIRI]} sisu (võimalik, et kirjutaja mainib ka lisaks loetud tekste või annab tagasisidet loengupidajale).}
&
\PromptCell{\# Your role and general instruction\par
You are the AI teaching assistant for the course ``Digital Competence and Artificial Intelligence''. Your task is to give feedback on the student's self-reflection essay presented below (it begins after the three \texttt{"""} marks). Take this feedback guide into account, consider all of its aspects, and on that basis synthesize brief and concrete feedback for the student (one polished, compact paragraph).\par
The essay's topic is the content of \texttt{[LECTURER]}'s lecture \texttt{[TITLE]} (the writer may also mention additional texts they have read or give feedback to the lecturer).}
\\
\PromptCell{\# Tagasisidestamise juhend: mida tagasisides mainida\par
\#\# Vorm ja keel, kontrolli kas:\par
- tekst on pikkuse suurusjärgus 300--350 sõna? (maini ainult siis, kui on oluliselt lühem või pikem)\par
- kirjutaja keelekasutus on üldiselt akadeemiline, selge, ja väldib üldsõnalisust?\par
- tekstil on hea ja loogiline struktuur?}
&
\PromptCell{\# Feedback guide: what to mention in the feedback\par
\#\# Form and language, check whether:\par
- is the text roughly 300--350 words long? (mention this only if it is substantially shorter or longer)\par
- is the writer's language generally academic and clear, and does it avoid vague or overly general wording?\par
- does the text have a good and logical structure?}
\\
\PromptCell{\#\# Sisu ja ülesande täitmine, analüüsi:\par
- Kas on vastatud nendele kolmele küsimusele: \texttt{[3 KÜSIMUST]}\par
- kas essees on lühike ülevaade loengust ja mis mõtted kirjutajal tekkisid?\par
- aga kas essees väljendub lisaks kirjeldusele ka kuidas õppimine toimus ja kas/miks see kirjutaja jaoks oluline oli (mõtle tagasisidestades Ash\&Clayton DEAL mudelile, aga seda nimeliselt tsiteerida pole vaja)\par
- kas kirjutaja on loonud vähemalt ühe isikliku seose: loengu sisu varasemate teadmiste või väärtuste või hoiakute kontekstis; ja kas kirjutaja mainib, kas/kuidas õpitut saaks kasutada, kas on midagi mida teisiti teha või mõelda? (vastus võib ka olla, et hoiakud ei muutunud või uusi teadmisi ei saanud)\par
- ja kokkuvõtteks, kas kirjutaja piirdub vaid kirjeldamisega või liigub vähemalt ühe sammu võrra sügavamale (nt seletab, küsib, analüüsib, seob laiemalt mõne teksti või ühiskondliku nähtusega). Analüüsi, mis tasemel (Hatton\&Smith 4 tasandi süsteemis) refleksioon asub, ja suuna vajadusel järgmine kord minema sügavamale (sel juhul ütle ka lühidalt kuidas seda teha; väldi üldsõnalist ``lihtsalt ole sügavam'' soovitust).}
&
\PromptCell{\#\# Content and task completion, analyze whether:\par
- have these three questions been answered: \texttt{[3 QUESTIONS]}?\par
- does the essay contain a brief overview of the lecture and the thoughts that arose for the writer?\par
- in addition to description, does the essay also express how learning took place and whether/why it was important to the writer? (When giving feedback, think of the Ash\&Clayton DEAL model, but there is no need to cite it by name.)\par
- has the writer made at least one personal connection, placing the lecture content in the context of prior knowledge, values, or attitudes; and does the writer mention whether/how what was learned could be used, or whether there is anything to do or think differently? (The answer may also be that attitudes did not change or no new knowledge was gained.)\par
- finally, does the writer remain at description or move at least one step deeper (for example, explain, question, analyze, or connect more broadly to another text or social phenomenon)? Analyze the level of reflection within the Hatton\&Smith four-level system and, if needed, guide the writer to go deeper next time (in that case, also say briefly how to do so; avoid vague advice such as ``just be deeper'').}
\\
\PromptCell{\#\# Tagasiside toon ja vorm:\par
- ole toetav aga konstruktiivelt kriitiline! kui midagi on puudu või pealiskaudne vms probleem, siis ütle see konkreetselt välja!\par
- too välja, millised reflektsiooni aspektid on täidetud hästi ja milliseid (kui on puudulikud) võiks edaspidi arendada!\par
- seosta soovitused kirjutaja enda tekstis toodud mõtetega!\par
- soovitusi andes väldi üldsõnalisust! ole väga konkreetne ja ära raiska sõnu! tagasiside peaks olema üks lõik!\par
- lõpus esita avatud küsimus (nt ``Edasi mõtlemiseks'') või mõtisklus (nt ``Huvitav, miks''), mis julgustab kirjutajat mingil teemal või suunal edasi mõtlema (aga ära esita sellist küsimust, millele kirjutaja on juba tegelikult vastanud).}
&
\PromptCell{\#\# Tone and form of the feedback:\par
- be supportive but constructively critical; if something is missing, superficial, or otherwise problematic, say so explicitly\par
- point out which aspects of reflection are done well and which ones, if deficient, could be developed further\par
- connect the suggestions to the ideas presented in the writer's own text\par
- avoid vague or overly general suggestions; be very concrete and do not waste words; the feedback should be one paragraph\par
- at the end, present an open question (for example, ``For further thought'') or a reflection (for example, ``Interesting why ...'') that encourages the writer to think further about a topic or in a particular direction (but do not ask a question that the writer has already actually answered)}
\\
\PromptCell{\# Anna nüüd tagasisidet sellele esseele:\par
\texttt{"""}\par
\texttt{[lisa siia oma tekst]}}
&
\PromptCell{\# Now provide feedback on this essay:\par
\texttt{"""}\par
\texttt{[insert the essay text here]}}
\end{longtable}

\section{Task-specific LLM annotation prompts}

The three annotation prompts below are reproduced in the English form used for the structured output API runs.
Each task was prompted separately so that the model saw only the text relevant to that coding decision. 
We used GPT-5.4 via its Responses API and structured outputs system, setting temperature to 0 and reasoning effort to none, to enable replicable outputs.

\subsection{Student response: appraisal stance and theme}

\SinglePrompt{Analyze this student's reflection on AI feedback to their reflective essay.\par
Text is in Estonian.\par
Return JSON only.\par
The student was asked to: ``Summarize your view of the feedback in 1--2 sentences. Was it substantive and useful, or were there mistakes or other problems?'' The student knew the feedback was AI generated.\par
Because of parsing noise, the input may also contain essay text or feedback text. Focus only on the student's own reaction to the feedback.\par
Output the most fitting or primary description for these 2 variables:\par
topic:\par
- actionable\_support: student says AI provided clear, usable guidance that can be directly applied to revision or future writing\par
- generic\_repetitive\_feedback: student says AI feedback feels template-like and repetitive, relying on abstract prompts while lacking individualized or specific critique\par
- constraint\_mismatch: student says AI requests deeper analysis or expansion that seems infeasible within word limits, task scope, time, or personal boundaries\par
- overly\_nice\_tone: student says AI is overly friendly or flattering and avoids sharp criticism\par
- misread\_or\_misunderstood: student says AI misinterpreted the text, overlooked stated content, or attributed claims the student did not make\par
- reliability\_variability: student says repeated uses or different AI tools give inconsistent evaluations, lowering trust\par
- other: fallback if none fit best, including technical constraints, word-count issues, sharing problems, language novelty, platform comparisons, or irrelevant commentary\par
sentiment:\par
- helpful: helpful or useful\par
- unhelpful: critical or negative about AI\par
- mixed: neutral, indifferent, or a mix of positive and negative\par
\par
Choose one primary topic and one sentiment.}

\subsection{Essay depth: Hatton--Smith level}

\SinglePrompt{Analyze this student's reflective essay.\par
Text is in Estonian.\par
Return JSON only.\par
The student was told: ``Write a reflective essay based on the lecture and optionally the extra readings. About 300--350 words, compact, clear, and well-structured. Briefly describe what happened in the lecture and what you learned. Include a reaction, question, or feeling. Connect the content to your prior knowledge, values, or attitudes. Link it to a broader text, theory, or social phenomenon. Explain whether and how you may use what you learned in the future, or what may change in your future thinking or action.''\par
Because of parsing noise, the input may also contain feedback and a later reaction to that feedback. Focus on the essay and output this value:\par
Hatton\_Smith\_level:\par
- descriptive: records events, content, or opinions only; little or no reflection; no real why, alternatives, or analysis\par
- descriptive\_reflection: mostly descriptive, but gives reasons or some reflection from at least one viewpoint; limited analysis and limited consideration of alternatives\par
- dialogic\_reflection: steps back and works through the experience; questions own views, explores alternatives, compares explanations, and links perspectives or factors\par
- critical\_reflection: examines assumptions and frame of reference; considers multiple perspectives and consequences for self and others; recognizes how context shapes interpretation\par
\par
Choose the highest level clearly supported. If uncertain, choose the lower level.}

\subsection{Feedback specificity: essay-groundedness prompt}

\SinglePrompt{Analyze how generic versus specific the feedback is. Text is in Estonian.\par
You will receive: - the student's essay; - the feedback on that essay.\par
Code the feedback only. Use the essay only as context for judging whether the feedback is clearly grounded in this particular essay. Do not judge correctness, overall quality, or pedagogical value. Do not score by length alone; short feedback can still be specific.\par
Return JSON only. Variable: specificity\_5pt\par
1 = very generic: could fit almost any essay; mostly broad praise, vague advice, or template-like comments; little or no clear reference to this essay.\par
2 = mostly generic: some limited anchoring in the essay topic, but still largely broad or reusable; few concrete references.\par
3 = mixed / moderately specific: both generic comments and some essay-specific remarks; at least one part clearly refers to this essay, but much remains broad.\par
4 = mostly specific: clearly grounded in this essay; refers to particular points, themes, structure, arguments, or omissions; advice is tailored rather than boilerplate.\par
5 = highly specific: strongly tied to this exact essay throughout; multiple concrete references to its actual content, argumentation, structure, examples, or missing elements; would not fit many other essays without major changes.\par
Interpret specificity by how clearly the feedback is anchored in the actual content of the essay and how far it avoids boilerplate or reusable comments. Implementation note: Only run this on rows where essay, feedback, and student reaction are all non-empty, and where \texttt{suspected\_parse\_failure} is false. Even though the model only sees essay + feedback, requiring all three parts reduces segmentation-error cases.}

\section{Supplementary model specifications and coefficients}

All mixed models specified a random intercept for student.
Linear and logistic mixed models were fit with the \texttt{bobyqa} optimizer and \texttt{maxfun = 100000}; the ordinal model used \texttt{ordinal::clmm} with a logit link and \texttt{nAGQ = 1}.
Tables~\ref{tab:appendix-model-spec}--\ref{tab:appendix-specificity} report fixed-effect estimates and standard errors without coefficient-level p-values; model names appear as row-group headings and ordinal cut-points are omitted.
The two p-values reported in the main paper come from likelihood-ratio comparisons of nested models and are reported once in Table~\ref{tab:appendix-lrt}.
Each full and reduced prompt-version model used 2988 observations from 283 students.
All four fits converged without optimizer messages and were non-singular; student random-intercept variances were 56.5 and 54.2 for the full and reduced low-specificity models, and 1.70 and 1.71 for the corresponding high-specificity models.
Refitting these models with Nelder--Mead reproduced the estimates and likelihood-ratio tests to the reported precision.

\begin{longtable}{P{0.20\textwidth}P{0.14\textwidth}P{0.40\textwidth}P{0.10\textwidth}}
\caption{Model specification summary for the supplementary coefficient tables.}\label{tab:appendix-model-spec}\\
\toprule
\textbf{Outcome} & \textbf{Family} & \textbf{Fixed effects} & \textbf{Student RI} \\
\midrule
\endfirsthead
\toprule
\textbf{Outcome} & \textbf{Family} & \textbf{Fixed effects} & \textbf{Student RI} \\
\midrule
\endhead
\bottomrule
\endfoot
Next observed grade & LMM & helpfulness + current grade + homework number + prompt version & yes \\
Same-homework grade & LMM & helpfulness + homework number + prompt version & yes \\
Grade change to next observed homework & LMM & helpfulness + current grade + homework number + prompt version & yes \\
Final essay outcome & LM & helpful share + unhelpful share + mean homework grade + number of coded reflections & no \\
Next observed reflection depth & ordinal mixed logit & helpfulness + current depth + homework number + prompt version & yes \\
Low specificity (1--3) & logistic mixed logit & homework number + prompt version & yes \\
High specificity (5) & logistic mixed logit & homework number + prompt version & yes \\
Specificity by essay context & LMM & essay depth + current grade + homework number + prompt version & yes \\
Helpful appraisal from specificity & logistic mixed logit & specificity + homework number + prompt version & yes \\
Genericity complaint from specificity & logistic mixed logit & specificity + homework number + prompt version & yes \\
\end{longtable}

\begingroup
\small
\setlength{\tabcolsep}{5pt}
\begin{longtable}{P{0.68\textwidth}R{0.12\textwidth}R{0.12\textwidth}}
\caption{Downstream and grade-related coefficients retained outside the main paper. Controls are documented in Table~\ref{tab:appendix-model-spec}.}\label{tab:appendix-downstream}\\
\toprule
\textbf{Fixed effect} & \textbf{$b$} & \textbf{$SE$} \\
\midrule
\endfirsthead
\toprule
\textbf{Fixed effect} & \textbf{$b$} & \textbf{$SE$} \\
\midrule
\endhead
\bottomrule
\endfoot
\multicolumn{3}{l}{\textit{Next observed homework grade}} \\*
Intercept & 2.260 & 0.054 \\*
Helpful (reference: mixed) & -0.028 & 0.017 \\*
Unhelpful (reference: mixed) & 0.017 & 0.028 \\*
Current grade & 0.149 & 0.019 \\*
Homework number & 0.005 & 0.004 \\*
Updated prompt (reference: initial) & 0.123 & 0.027 \\
\addlinespace[0.5ex]
\multicolumn{3}{l}{\textit{Same-homework grade}} \\*
Intercept & 2.633 & 0.021 \\*
Helpful (reference: mixed) & -0.013 & 0.017 \\*
Unhelpful (reference: mixed) & -0.012 & 0.027 \\*
Homework number & -0.002 & 0.003 \\*
Updated prompt (reference: initial) & 0.221 & 0.024 \\
\addlinespace[0.5ex]
\multicolumn{3}{l}{\textit{Grade change to next observed homework}} \\*
Intercept & 2.260 & 0.054 \\*
Helpful (reference: mixed) & -0.028 & 0.017 \\*
Unhelpful (reference: mixed) & 0.017 & 0.028 \\*
Current grade & -0.851 & 0.019 \\*
Homework number & 0.005 & 0.004 \\*
Updated prompt (reference: initial) & 0.123 & 0.027 \\
\addlinespace[0.5ex]
\multicolumn{3}{l}{\textit{Final essay outcome}} \\*
Intercept & 2.376 & 0.107 \\*
Helpful share & 0.060 & 0.034 \\*
Unhelpful share & 0.069 & 0.072 \\*
Mean homework grade & 0.185 & 0.038 \\*
Number of coded responses & -0.001 & 0.003 \\
\addlinespace[0.5ex]
\multicolumn{3}{l}{\textit{Next observed reflection depth}} \\*
Helpful (reference: mixed) & 0.204 & 0.221 \\*
Unhelpful (reference: mixed) & 0.269 & 0.363 \\*
Current depth & 0.285 & 0.257 \\*
Homework number & 0.014 & 0.050 \\*
Updated prompt (reference: initial) & -0.507 & 0.346 \\
\end{longtable}
\endgroup

\begingroup
\small
\setlength{\tabcolsep}{5pt}
\begin{longtable}{P{0.68\textwidth}R{0.12\textwidth}R{0.12\textwidth}}
\caption{Specificity, context, and response-model coefficients. Controls are documented in Table~\ref{tab:appendix-model-spec}.}\label{tab:appendix-specificity}\\
\toprule
\textbf{Fixed effect} & \textbf{$b$} & \textbf{$SE$} \\
\midrule
\endfirsthead
\toprule
\textbf{Fixed effect} & \textbf{$b$} & \textbf{$SE$} \\
\midrule
\endhead
\bottomrule
\endfoot
\multicolumn{3}{l}{\textit{Low specificity (scores 1--3)}} \\*
Intercept & -9.934 & 1.200 \\*
Homework number & 0.164 & 0.150 \\*
Updated prompt (reference: initial) & -3.210 & 1.228 \\
\addlinespace[0.5ex]
\multicolumn{3}{l}{\textit{High specificity (score 5)}} \\*
Intercept & -5.059 & 0.300 \\*
Homework number & 0.138 & 0.031 \\*
Updated prompt (reference: initial) & 1.090 & 0.330 \\
\addlinespace[0.5ex]
\multicolumn{3}{l}{\textit{Specificity by essay context}} \\*
Intercept & 3.820 & 0.060 \\*
Depth level (1--4) & 0.051 & 0.022 \\*
Current grade & 0.015 & 0.015 \\*
Homework number & 0.010 & 0.003 \\*
Updated prompt (reference: initial) & 0.041 & 0.021 \\
\addlinespace[0.5ex]
\multicolumn{3}{l}{\textit{Helpful student response}} \\*
Intercept & 0.198 & 0.582 \\*
Specificity (1--5) & 0.181 & 0.143 \\*
Homework number & -0.057 & 0.020 \\*
Updated prompt (reference: initial) & -0.216 & 0.157 \\
\addlinespace[0.5ex]
\multicolumn{3}{l}{\textit{Genericity complaint in student response}} \\*
Intercept & -2.258 & 1.034 \\*
Specificity (1--5) & -0.498 & 0.252 \\*
Homework number & 0.013 & 0.034 \\*
Updated prompt (reference: initial) & 0.876 & 0.299 \\
\end{longtable}
\endgroup

\begin{longtable}{P{0.20\textwidth}P{0.38\textwidth}P{0.08\textwidth}P{0.06\textwidth}P{0.07\textwidth}}
\caption{Likelihood-ratio tests corresponding to the two mixed-model p-values reported in the main paper. The reduced model omits prompt version but retains homework number and the student random intercept.}\label{tab:appendix-lrt}\\
\toprule
\textbf{Outcome} & \textbf{Comparison} & \textbf{$\chi^2$} & \textbf{$df$} & \textbf{$p$} \\
\midrule
\endfirsthead
\toprule
\textbf{Outcome} & \textbf{Comparison} & \textbf{$\chi^2$} & \textbf{$df$} & \textbf{$p$} \\
\midrule
\endhead
\bottomrule
\endfoot
Low specificity (1--3) & Full versus reduced model & 8.77 & 1 & .003 \\
High specificity (5) & Full versus reduced model & 11.63 & 1 & $<.001$ \\
\end{longtable}

\else
\paragraph{Online appendix.} Omitted for anonymous review; will provide anonymized workflow metadata, prompt versions, translations, and LLM annotation prompts.
\bibliographystyle{splncs04}
\bibliography{lit}
\fi

\end{document}